# Redundancy Generation in University-Industry-Government Relations: The Triple Helix Modeled, Measured, and Simulated

Inga A. Ivanova [1]* & Loet Leydesdorff [2]

**Abstract**

A Triple Helix (TH) of bi- and trilateral relations among universities, industries, and governments can be considered as an ecosystem in which uncertainty can be reduced auto-catalytically. The *correlations* among the *distributions of relations* span a vector space in which two vectors ($P$ and $Q$) represent "sending" and "receiving," respectively. These vectors can also be understood in terms of the generation versus reduction of uncertainty in the communication field that results from interactions among the three (bi-lateral) communication channels. We specify a set of Lotka-Volterra equations between the vectors that can be solved. Redundancy generation can then be simulated and the results can be decomposed in terms of the TH components. Among other things, we show that the strength and frequency of the relations are independent parameters. Different components in terms of frequencies in triple-helix systems can also be distinguished and interpreted using Fourier analysis of the empirical time-series. The case of co-authorship relations in Japan is analyzed as an empirical example; but "triple contingencies" in an ecosystem of relations can also be considered more generally as a model for redundancy generation by providing meaning to the (Shannon-type) information in inter-human communications.

[1] Far Eastern Federal University, School of Business and Public Administration, 8 Sukhanova St., Vladivostok, 690091, Russia; inga.iva@mail.ru; * corresponding author.
[2] Amsterdam School of Communication Research (ASCoR), University of Amsterdam, Kloveniersburgwal 48, 1012 CX Amsterdam, The Netherlands; loet@leydesdorff.net .

# 1. Introduction

Leydesdorff & Ivanova (in press) argue that mutual information in three (or more) dimensions (McGill, 1954) does not measure Shannon-type information, but "mutual redundancy." As against Shannon-type information that can only be positive, mutual redundancy is a signed information measure (Yeung, 2008, pp. 59f.). Negative values indicate a reduction of uncertainty. The Shannon-type (and thus positive) information generated in interactions among more than two sources of information, however, can be approximated using Krippendorff's (1980; 2009a and b) formulas for interaction information ($I_{ABC \to AB:AC:BC}$).

"Mutual redundancy" is generated when differently positioned sources of information overlap in terms of the distributions that participate in the communication. The overlaps among the communication channels result in an additional communication field that can cause feedback on the relations depending on the positions of the sources of variation. When the agents are changing in terms of variable relations, the communication field also changes because the new relations can be expected to mean other things in the different directions (and for differently positioned agents).

Thus, two layers can be distinguished analytically: at the basis, the layer of mutual *relations,* and thereupon the layer of latent structures based on *correlations* in the data, when more than two sources are involved. These structures, however, do not relate; they overlap and permeate one another—or, in other words, radiate as fields into one another's spheres of influence. These structures change with the relations operating over time. The relations provide variation, and the structures operate as selection mechanisms and therefore can be expected to change at a rate lower than that of the variation.

In other words, the structural layer contains the latent dimensions of changes observable in terms of network relations. The changes at this next-order level are not (Shannon-type) information exchanges as in the lower layer of relations, but rather a structuration of these exchanges. The exchanges in these relations can also be considered as instantiations of the structures operating in historical configurations (Giddens, 1979). However, one can expect the structures to impact as feedback on the information exchanges. The latent factors can, for example, lead to spuriousness in the correlations.

In the case of three (or more) sources of variation, the net result between information generation in the relations and redundancy generation with respect to differently positioned structures can be unequal to zero (Leydesdorff & Ivanova, in press). In studies of the Triple Helix of university-industry-government relations this positional dynamic at the structural level has been called a "communication overlay" by Etzkowitz & Leydesdorff (2000) or a "hyper-cycle of communications" (Leydesdorff, 1994; 2006). In this study, we consider this hyper-cycle as a communication field resulting from the interaction of three (or more) communication channels.

An example of the possible feedback of the communication field as a latent construct is provided by the situation of a child asking a question of one of two parents while being able to predict the answer of the other parent. The latent structure of the marriage functions in the background and reduces the uncertainty that prevails in the configuration. If the marriage falls apart in a divorce, this reduction of uncertainty at the systems level may again disappear or even be reversed.

We model the latent communication field *W* in terms of two three- (or more-)dimensional vectors: *P* and *Q*. In a model of reflexive communications, one needs two vectors because Shannon-type information will necessarily be generated with the arrow of time while redundancies can be the result of shared meanings provided with the perspective of hindsight. For reasons of presentation, we limit our discussion to the three-dimensional case.

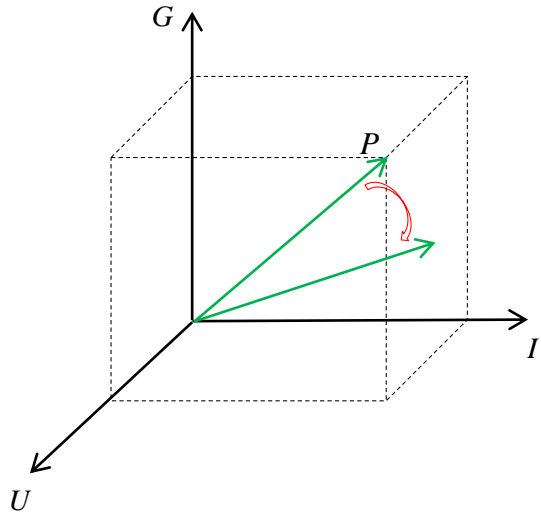

**Figure 1:** *P* (and *Q*) are vectors in the three-dimensional space of *U*, *I*, and *G*.

Figure 1 shows the vector *P* and its possible rotation in the vector space of university-industry-government (*UIG*) relations given states of the system in terms of specific sets of relations. Such a state can also be summarized in terms of two vectors *P* (for the sending) and *Q* (for the reflexive receiving) with different values in the three dimensions. The two three-dimensional vectors include all interaction terms, such as in bilateral and trilateral relations.

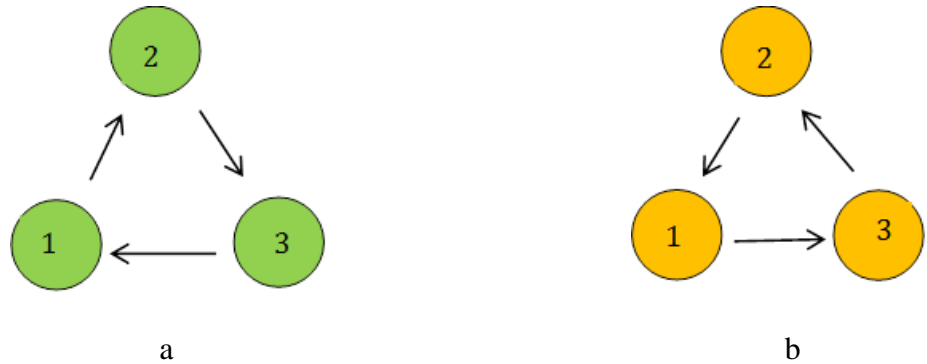

**Figure 2:** Circulation and feedback in cycles in both directions (*P* and *Q*).

The second vector is generated in a complex system: the feedback cycles among the helices can be counter-clockwise as well as clockwise as the partners are continuously able to feedback and feed forward upon one another. The two directions are distinguished analytically in Figure 2. We elaborate on Ulanowicz' (2009a, at p. 1888, Figure 3)[3] schematic depiction of auto-catalysis in the left-side picture (a) of Figure 2 (see also Ulanowicz, 2008 and 2009b). Padgett & Powell (2012, at p. 55) have noted that auto-catalysis can be considered as another term for *autopoiesis* (Maturana & Varela, 1980) or self-organization (Luhmann, 1986). As against the biological model, we add a reflexive dynamic in Figure 2b that remains emergent as an "overlay," but continuously absorbs reflexively the uncertainty generated in the underlying dynamics.

The two cycles in Figure 2 provide us phenotypically with a net balance between two dynamics. One can also consider this balance as a trade-off between evolutionary self-organization and historical organization, (Leydesdorff, 2010) or, in other words, between recursion on a previous state along the historical axis as opposed to meaning provided to the events from the perspective of hindsight (Dubois, 1998). The self-organization of a social system adds the exchanges of expectations in an overlay of communcations to the relational exchanges of (Shannon-type) information (Leydesdorff, 2011, 2012; Luhmann, 1986 and 1995). The consequent overlay or communication field remains fractal, fragmented, and fragile because it feeds back on its instantiations at specific moments in time. These instantiations can also be considered as the retention mechanism of a complex system of fluxes.

---

[3] Ulanowicz (2009a, at p. 1888) provides the following illustration: 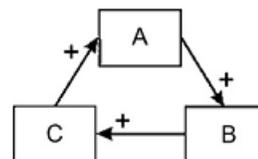 . However, one can add minus signs to the tails of the arrows in this figure, and thus obtain the trade-off between evolutionary auto-catalysis ("self-organization") versus historical organization as in Figures 2a and 2b.

The empirical question is which tendency—historical integration into (networked) organizations at specific moments of time or evolutionary differentiation coordinated by self-organization over time—can be expected to prevail? If historical organization prevails, the system is more hierarchical than decentralized, whereas self-organization can lead to emerging systemness with synergies among distributions—and therefore local reductions of uncertainty. In other words, the vectors *P* and *Q* can be coupled differently, and one cannot expect the cycles to be synchronized. We shall study parameters such as the coupling coefficient (*g*) among the helices and introduce stochastic behavior into the model in terms of fuzziness in the intervals between the two rotations.

The reflexive overlaps among the structures—the correlations among the lower-order relations—can lead to redundancy. In that case, the net result can be a reduction or increase of the prevailing uncertainty in an empirical system (e.g., Leydesdorff & Sun, 2009; Ki-seok *et al.*, 2012; Strand & Leydesdorff, 2013; Leydesdorff & Strand, in press). The equation for the mutual redundancy in terms of Shannon-type uncertainties expresses this possibility of positive versus negative values as follows (e.g., Abramson, 1963, at pp. 129 ff.; Leydesdorff & Ivanova, in press; McGill, 1954; Yeung, 2008, at pp. 59f.):

$$R_{123} = H_1 + H_2 + H_3 - H_{12} - H_{13} - H_{23} + H_{123} \tag{1}$$

Unlike vectors (such as *P*, *Q*, and *W*), information measures are scalars. In order to relate the vectors to these scalars, we propose to use scalar products by squaring the vectors as follows:

$$W^*W = P^*P - Q^*Q \tag{2}$$

It follows that:

$$R_{123} \sim W^2 \tag{3}$$

The precise functional dependence between redundancy *R* and the communication field *W* can be left unspecified for the moment.[4] What is crucial is that both *R* and *W* indicate a structural

---

[4] In the case of weak coupling among the three dimensions, one can use perturbation theory for an approximate solution. By writing the solution in terms of a formal power or perturbation series, one can formulate for relatively small parameters $W^2$ (<< 1), as follows:

$$R = \alpha_0 + \alpha_1 W^2 + \alpha_2 (W^2)^2 + \cdots \tag{4a}$$

Since raising the power of $W^2$ further leads to increasingly smaller values, one can approximate the perturbation solution by using the first two terms:

$$R = \alpha_0 + \alpha_1 W^2$$

difference between two dynamics (Krippendorff, 2009b), namely: a forward dynamic with the historical axis of time generating Shannon-type entropy, and a reflexive dynamic that can operate from the perspective of hindsight.

In the Triple-Helix case, the *two* vectors *P* and *Q* operate as *three*-dimensional (*UIG*) selection environments upon each other. Selection environments operating upon one another can be modeled using Lotka-Volterra equations; but given the three dimensions, we have to develop these equations for vectors instead of scalars (Appendix A). Because we are interested in the selection terms, and one may assume that variation takes place stochastically in the (lower-order) network of relations—and thus averages to zero—the equations can be solved analytically (Appendix B). The three components of the vectors can then be simulated, squared, and the value of the prevailing redundancy determined as a result of the subtraction (Eq. 2).

The solution of the equations will enable us to consider TH dynamics also as Fourier series. We use Fourier-series analysis for the decomposition of an empirical study of three-dimensional redundancies in the four-dimensional case of national and international university-industry-government co-authorship relations in Japan (Leydesdorff & Sun, 2009). The four-dimensional case enables us to compare four three-dimensional ones. The relative importance of the different cycles (e.g., university-industry-government *versus* university-industry-international) can be specified and each three-dimensional (sub)system can be analyzed further in terms of its composing frequencies.

In summary, we make three steps in this study: first, we derive the model mathematically by extending Lotka-Volterra equations to the case of three selection environments operating upon one another. Secondly, we run the model for variations in some relevant parameters (e.g., the coupling coefficient). The simulations are provided in an Excel workbook (at http://www.leydesdorff.net/redundancy/figures.xlsx ) so that the reader is enabled to extend on this model and/or to vary parameters. Finally, we show how one can decompose previously published empirical results in terms of this model using Fourier analysis of the time series. The application remains exemplary given the limitations of this data.

## 2. The Triple Helix overlay as a communication field

In terms of structures, the three interacting agents in TH relations (university, industry, government) shape selection environments for one another while interacting. Whereas the relational network can be considered as specific integrations in historical instances, the structural

---

or, disregarding the constant term, as:

$$R \sim W^2 \qquad (4b)$$

*Q.e.d.*

environments provide different selection mechanisms. The three selection environments operate in terms of generating wealth (industry) and novelty (academia), or by providing governance (regulation and legislation).

The observable relations can be studied as variation using, for example, social network analysis. However, the functional selections require dynamic modeling using Lotka-Volterra equations that can first be formulated as follows:

$$\begin{cases} P_t = \alpha P - \beta P Q \\ Q_t = -\gamma Q + \delta Q P \end{cases} \quad (5)$$

In Eq. (5), however, $P$ and $Q$ are scalars. In our case of coupling among three or more environments, we propose to use the vector form of Equation (5) by defining $P$ and $Q$ as vectors with three components (Figure 1), and formulate as follows: $P = \begin{pmatrix} P_1 \\ P_2 \\ P_3 \end{pmatrix}; Q = \begin{pmatrix} Q_1 \\ Q_2 \\ Q_3 \end{pmatrix}$.

Eq. (5) can then be written as follows:

$$\begin{cases} P_t = \alpha P - \beta (P \times Q) \\ Q_t = -\gamma Q + \delta (Q \times P) \end{cases} \quad (6)$$

Coefficients $\alpha$ and $\gamma$ can be set equal to zero because these terms model the variation (in the relations) and not the mutual selections. This simplification allows us to solve the equations analytically (in Appendix B) so that the three components of the two vectors can be distinguished as such:

$$\begin{aligned} P_{1t} &= -2g[P_2 Q_3 - P_3 Q_2] \\ P_{2t} &= -2g[P_3 Q_1 - P_1 Q_3] \\ P_{3t} &= -2g[P_1 Q_2 - P_2 Q_1] \end{aligned} \quad (7)$$

$$\begin{aligned} Q_{1t} &= 2g[P_2 Q_3 - P_3 Q_2] \\ Q_{2t} &= 2g[P_3 Q_1 - P_1 Q_3] \\ Q_{3t} &= 2g[P_1 Q_2 - P_2 Q_1] \end{aligned} \quad (8)$$

The parameter $g$ is the coupling coefficient (which is derived in Appendix A). Using simulations, we show below, among other things, that the value of $g$ determines the frequencies in the helices.

Let us multiply the first equation of System (7) by $P_1$, the second by $P_2$, the third by $P_3$, and sum the three equations in order to obtain:

$$P_1 P_{1t} + P_2 P_{2t} + P_3 P_{3t} = 0 \tag{9}$$

Integration of this equation over time leads to an invariant:

$$P_1^2 + P_2^2 + P_3^2 = C_1 \tag{10}$$

Similarly, one can derive another invariant from System (8):

$$Q_1^2 + Q_2^2 + Q_3^2 = C_2 \tag{11}$$

$C_1$ can be considered an analogue to the sum of the negative terms or the redundancy in Eq. 1, and $C_2$ as the sum of the positive (Shannon-type) information terms. It follows that $C = C_2 - C_1$ is also an invariant. $C$ can be considered as $W^2$—the square of the communication field—in Eq. 4, as follows:

$$[C = C_2 - C_1] \sim [W^2 = P^2 - Q^2] \sim R \tag{12}$$

The interpretation of Eq. 12 in terms of information measures (e.g., bits) is allowed because the reasoning is hitherto dimensionless. The Systems (7) and (8) can also be considered as the clockwise *versus* counter-clockwise cycles of Figure 2. Alternatively, using the same formulas, one can interpret the cycles as spatially in the same direction, but one performed with the arrow of time—in terms of historical relations—and the other against the arrow of time. That is, by providing meaning to the uncertainty generated in the historical relations from the perspective of hindsight. The spatial and temporal dimensions {*x, t*} are equivalent at this level of abstraction.

In summary, we have shown that the three-dimensional vector *P* of university-industry-government relations in Figure 1 has a three-dimensional selection environment *Q*, and vice versa (Figure 2). This can also be seen as the alternation between sending and receiving among the three agents. One needs two (three-dimensional) vectors because the three agents can alternate in the directionality of their feedback, both spatially at each moment and over time in terms of taking turn or attributing meaning from a receptive perspective. As selection environments, the vectors operate upon each other using the modified Lotka-Volterra equations. The variation is provided by the observable relations, that is, forward action. Note that action is historical and can be stochastic, whereas selection is theoretically hypothesized (as evolutionary), and deterministic.

By squaring the two three-dimensional vectors $P_{[U,I,G]}$ and $Q_{[U,I,G]}$, we derived scalar invariants that can be subtracted. This subtraction can also be considered as the positive and negative terms in the mutual redundancy that results (Eq. 12). In the next section, we use the specification of the Triple Helix in terms of $P$ and $Q$ for the simulation of the value of this redundancy ($R \sim W^2 = P^2 - Q^2$). Our longer-term aim is to use empirical distributions to estimate these parameters and thus to use the model for the specification of an expectation. Here, we first develop the instruments for this longer-term objective.

## 3. Simulations

The equation systems (7) and (8) enable us to derive partial solutions for the three dimensions, as follows:

$$\begin{aligned} P_{1t} + Q_{1t} &= 0 \\ P_{2t} + Q_{2t} &= 0 \quad \text{or after integration:} \quad \begin{aligned} P_1 + Q_1 &= \alpha \\ P_2 + Q_2 &= \beta \\ P_3 + Q_3 &= \gamma \end{aligned} \end{aligned} \tag{13}$$

Substitution of $Q$ in terms of $P$ in system (13) results in:

$$\begin{aligned} P_{1t} &= cP_2 - bP_3 \\ P_{2t} &= aP_3 - cP_1 \\ P_{3t} &= bP_1 - aP_2 \end{aligned} \tag{14}$$

where $a = -2g\alpha$; $b = -2g\beta$; $c = -2g\gamma$. System (14) provides us with a well-known problem that has an analytical solution (Kamke, 1971). The derivation (in Appendix B) leads to a partial solution for each helix as the sum of a constant and of a time-dependent oscillation. Such a function can always be written as a composite of sine and cosine functions, as follows:

$$P_i = A_i + B_i \cos(rt) + D_i \sin(rt) \tag{15}$$

where $i = 1$, 2, or 3 (that is, $U$, $I$, or $G$), and $r$ is proportional to the coupling coefficient $g$ in Equations (7) and (8). A similar set of three equations of $Q_i$ can be derived for the system provided by Equation (8).

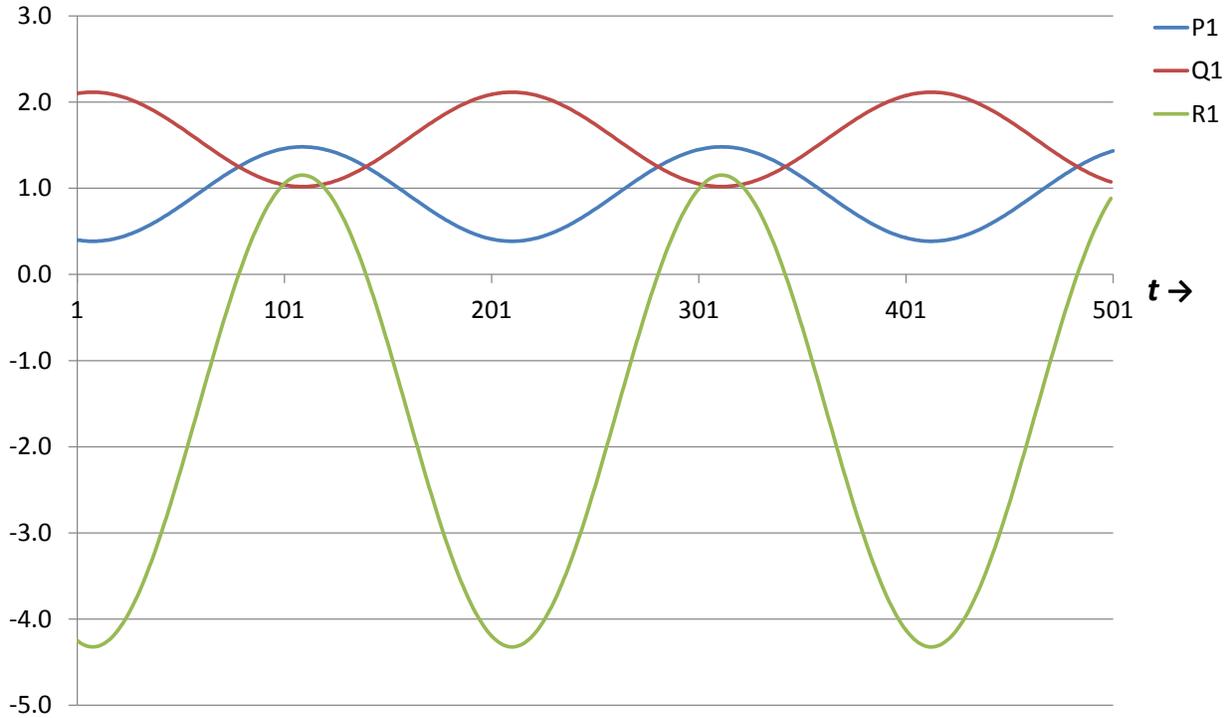

**Figure 3**: Longitudinal development of the two vectors $P$ and $Q$ for the first component ($U = 1$) and the consequent development of the contribution of this component to the redundancy $R_1$ (= $P_1^2 - Q_1^2$). The initial values of this simulation were: $P_1 = 0.4$, $P_2 = 1.4$, $P_3 = 0.7$, and $Q_1 = 2.1$, $Q_2 = 1.1$, and $Q_3 = 0.9$; the coupling coefficient $g = 0.2$. The y-axis uses absolute values (e.g., bits) and the x-axis the time steps of the simulation.

Figure 3 shows the simulation of Eq. 15 for $i = 1$—let's say for the academic partner in a Triple-Helix configuration. (The simulations are available for download as an Excel file at http://www.leydesdorff.net/redundancy/figures.xlsx.) One can first see that there is an alternation between the developments of the two vectors $P$ and $Q$. This can be interpreted as an alternation between the sending and receiving modes or—from a different perspective—as a difference between an anticipated state at $t = t + 1$ and the historical situation at $t = t$. (In a next section, we shall vary the interval between $P$ and $Q$ by adding noise to it.) The squared values of the two cycles can be subtracted (Eq. 2), and then the redundancy contribution for the first component $R_1$ (green line) is shown. The total redundancy generated by a Triple Helix system will, however, be an aggregate of the partial redundancies: $R_{123} = R_1 + R_2 + R_3 = \sum_{i=1}^{3}(P_i^2 - Q_i^2)$ because redundancy is additive.

## 3.1 *The effects of different coupling coefficients (g)*

Figure 4 shows the three solutions for *P* and *Q*, and the corresponding partial redundancies *R* for two different values of the coupling coefficients: $g = 0.2$ (on the left-side) and $g = 0.02$ (on the right side). For this illustration we used the following initial values for the three dimensions (*U, I, G*):

$$P_1(0) = 0.4 \qquad Q_1(0) = 0.2$$
$$P_2(0) = 1.4 \qquad Q_2(0) = 0.87 \qquad (16)$$
$$P_3(0) = 0.7 \qquad Q_3(0) = 0.9$$

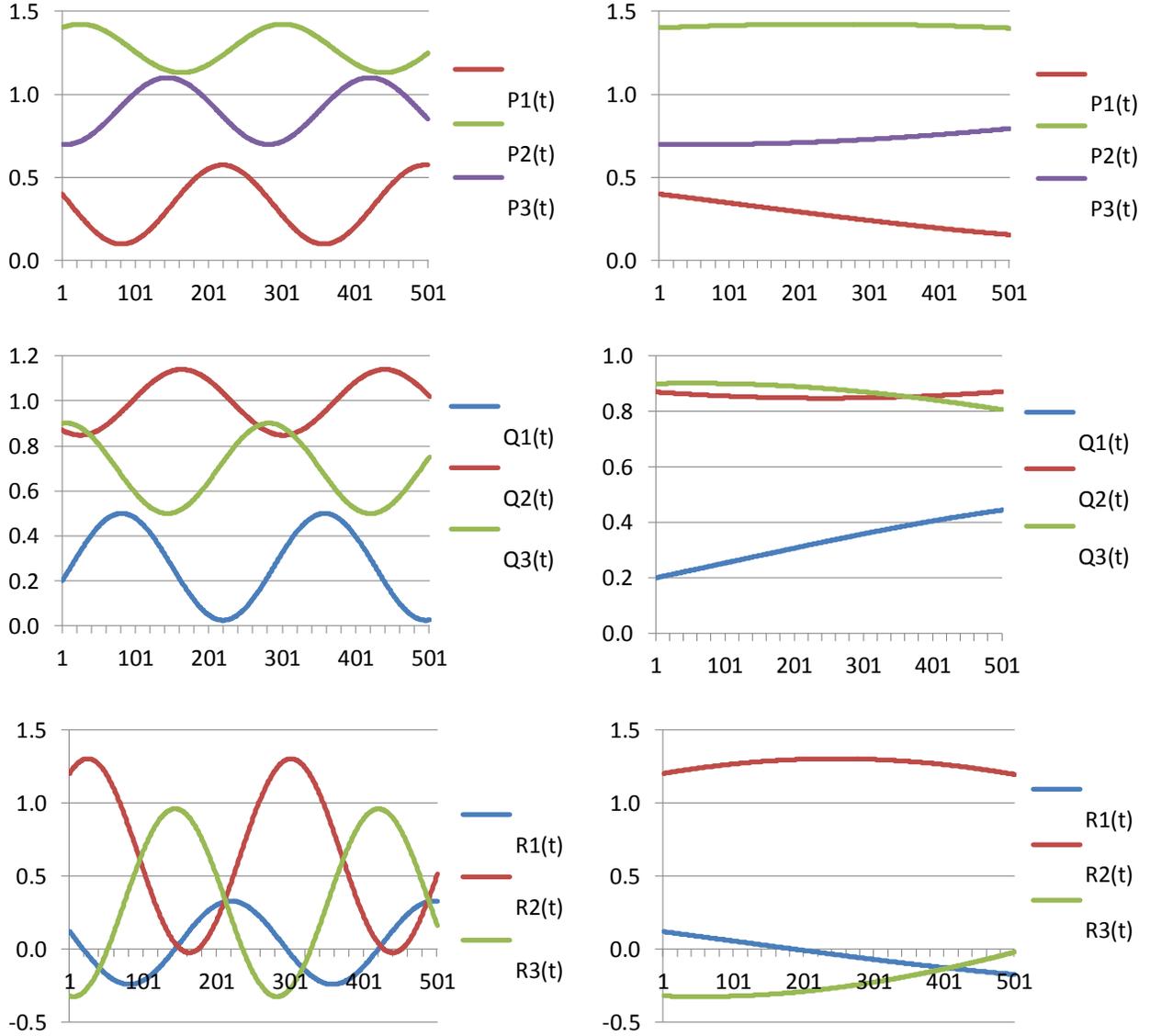

**Figure 4:** The effect of the coupling constant *g* on the time-dependence of the partial components *P*, *Q*, and *R*) in the three dimensions *U,I,G* for *g*=0.2 (on the left side), and *g*= 0.02 (on the right side). The *y*-axis uses absolute values (e.g., bits) and the *x*-axis the time steps of the simulation.

Note first that unlike the measurement of the mutual redundancy in terms of Eq. (1)—which leads to a *single* scalar value for the reduction of uncertainty as a systems property of a configuration—the *R*-values for the different strands of the TH ($R_1$, $R_2$, and $R_3$) can be distinguished in the simulation. Since redundancy is an additive function, one can formulate (see Eq. 3 above), as follows:

$$R = \sum_{i=1}^{3} R_i \sim \sum_{i=1}^{3} W_i^2 = \sum_{i=1}^{3}(P_i^2 - Q_i^2) \qquad (17)$$

In other words, one is able to specify how much redundancy the three subsystems (U, I, G) contribute to the total reduction of uncertainty and therefore the potential synergy ($R < 0$) in the TH configuration under study. We shall show below that the summation of the three components ($\sum_{i=1}^{3} R_i$) leads to a constant (that can also be considered as the scalar value resulting from the measurement).

Furthermore, Figure 4 shows that the periodical reshaping of the communication arrangement in the TH is dependent on the coupling coefficient *g*. The stronger the coupling, the higher the frequencies of the updates. If innovation capacity is dependent on the synergy in TH arrangements, one would thus be able to speed up innovation by increasing the coupling coefficients.[5] Note that this coefficient is specified in terms of correlations in the vector space, and not in terms of relations in the network.

Bringing the subsystems closer together in a multi-dimensional vector space can generate a niche in which the chances of innovation are enhanced (Biggiero, 1998, 2001; Kemp *et al.*, 1998). The relations between this vector space and the network space are non-linear, but the empirical data can be analyzed in terms of both topologies—the network graph and the vector space—because one is based on the variable relations and the other on correlations among the same variables as distributions (Leydesdorff, in press).

*3.2.    Adding stochastic fuzziness*

In the previous paragraph, we described the oscillations as smooth (sine and cosine) functions. However, redundancy is generated in an operation from the perspective of hindsight (because of the second law).[6] Dubois' (1998) model of anticipatory systems shows that a stochastic element can be involved structurally in providing meaning to the events based on expectations (Leydesdorff, 2010; Leydesdorff & Sander, 2009). We propose to introduce this stochastic element by assuming that the subtractions in Eq. 17 are not coherent, or—in other words—that the reaction time between *P* and *Q* (in Figure 3) can be disturbed stochastically.

In terms of the simulation, we introduce a random variable when modeling the relative phase shift among the two terms (*P* and *Q*) in Eq. 17. The value of this random phase shift varies within an interval that will be denoted as "the fuzzy interval." The length of this fuzzy interval (e.g., the reaction time at the receiving end) accounts for the degree of random fluctuations within limits. We show below that, for a fixed length of the fuzzy interval, the degree of randomization depends on the oscillation frequency induced by the coupling coefficient *g*: rapid oscillations can be more random compared with slowly oscillating cycles.

---

[5] In a next step, one would be able to differentiate among different coupling coefficients in the three (bilateral) communication channels.

[6] The second law of thermodynamics holds equally for probabilistic (Shannon) entropy, since $S = k_B H$ and $k_B$ is a constant (the Boltzmann constant). Because of the constant, the development of *S* over time is a function of the development of *H*, and *vice versa*.

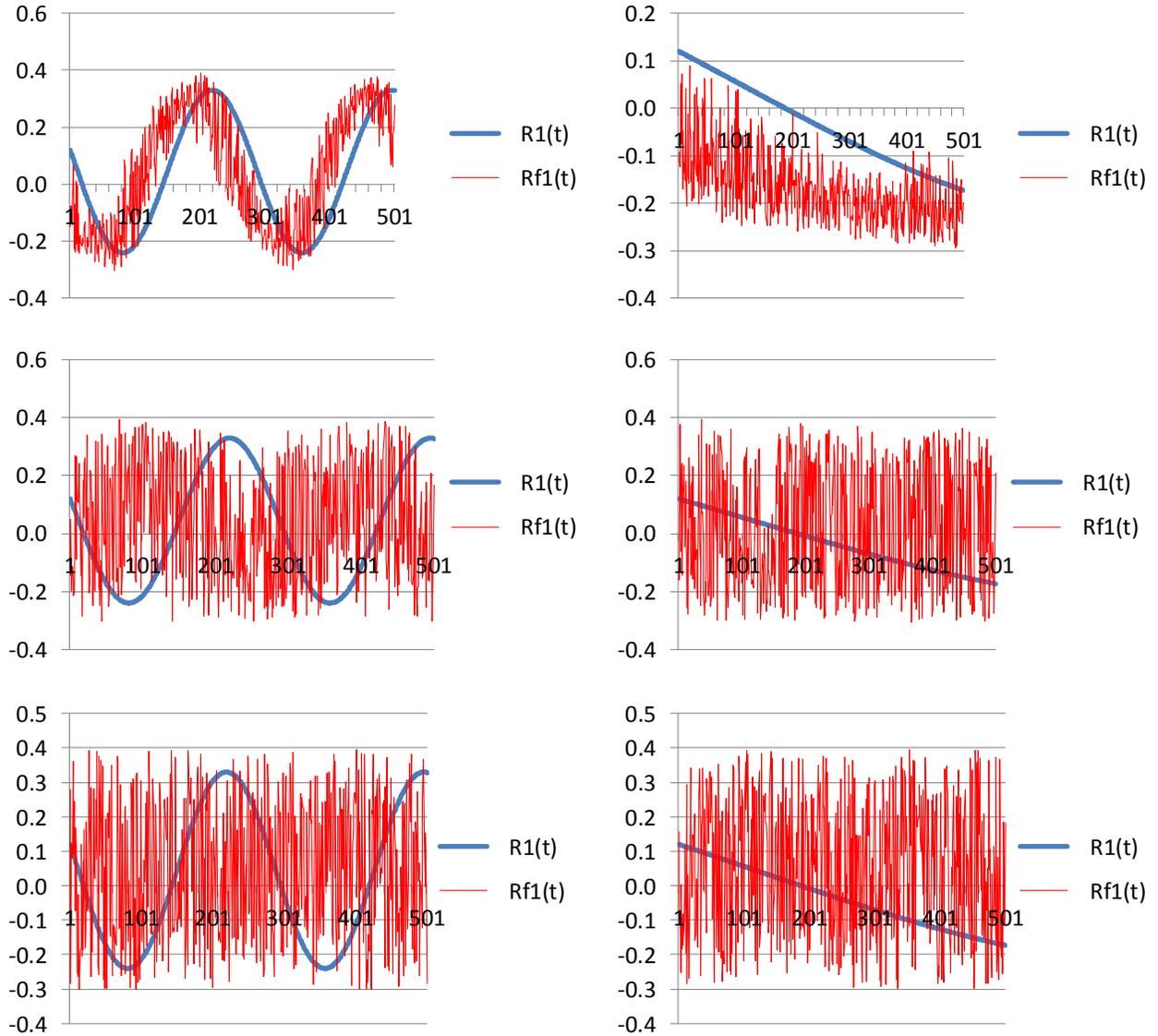

**Figure 5**: Effects of randomization on the partial redundancies generated with ($Rf_1$) or without ($R_1$) fuzzy intervals for two values of the coupling factor $g = 0.2$ (left column) and $g = 0.02$ (right column); the fuzzy intervals are $(0, \pi/2r)$ in the top row, $(0, 3\pi/2r)$ in the middle row, and $(0, 12\pi/2r)$ at the bottom.

Figure 5 shows the effects of randomization on the generation of redundancy in one of the TH strands ($R_1$). For analytical reasons, one can expect on the basis of Eq. 15 (above) that the length of fluctuation cannot exceed $3\pi/2r$ (Appendix B). This value is used in the middle row. The resulting figures (in the third row) show that a further increase of the fuzzy interval above this maximum value (of 3π/2r) does not change the pattern using a four times larger interval (0, 12π/2r). This maximum suggests a limitation to the anticipation: the next cycle can restructure the system to such an extent that the prediction becomes unreliable due to unintended consequences.

Below the maximum (in the first row of Figure 5), stochastic behavior depends on the length of the fuzzy interval, that is, on the system's reaction time upon receiving the information. By narrowing the fuzzy interval, one obtains more correlatable behavior between the blue and red curves in the top rows as compared to the middle ones. Such higher correlation—that is, shorter reaction times—corresponds with a more independent position for the three helix components, respectively.

The right column of Figure 5 shows that looser coupling (that is, a lower value of *g*) does not affect the stochastic patterns in the redundancies generated. In summary, these simulation results suggest that the strength of the interactions—that is, the extent to which the cycles of the helices can be preserved despite disturbances—and their response times—determined by the coupling coefficients *g*—are two distinct parameters.

*3.3.    Summation of the redundancies in the three components*

The effects of randomization on the summary redundancy, obtained via summation of the redundancies in all three dimensions, are shown in Figures 6 and 7, respectively.

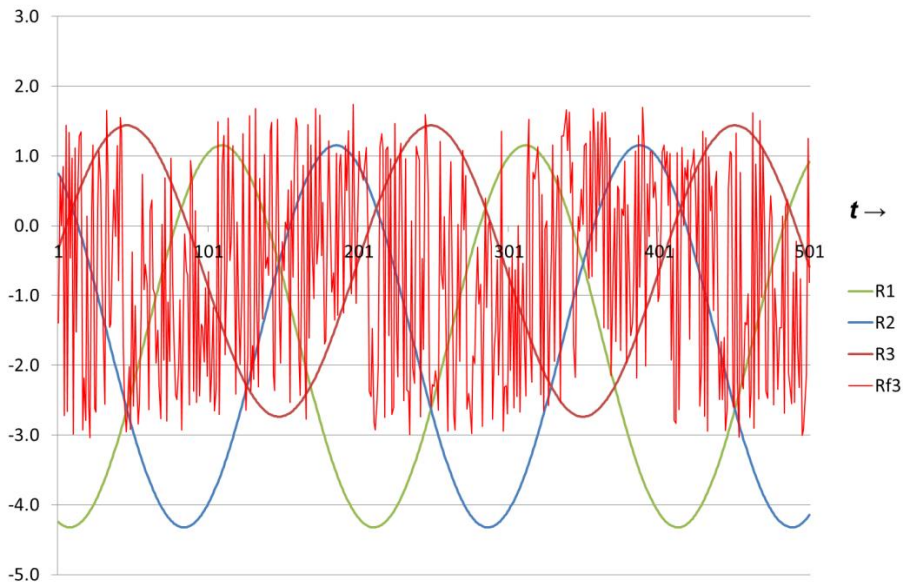

**Figure 6**: Three components in the generation of redundancy with noise in the fuzzy interval of (0, 3π/2r) added to the third component $Rf_3$: initial values as in Figure 3.

Figure 6 shows the three components of the redundancy in university-industry-government relations. (The green curve for $R_1$ is the same as in Figure 3.) We added (maximum) noise for the fuzzy interval (0, 3π/2r) to the third component ($Rf_3$) as an illustration.

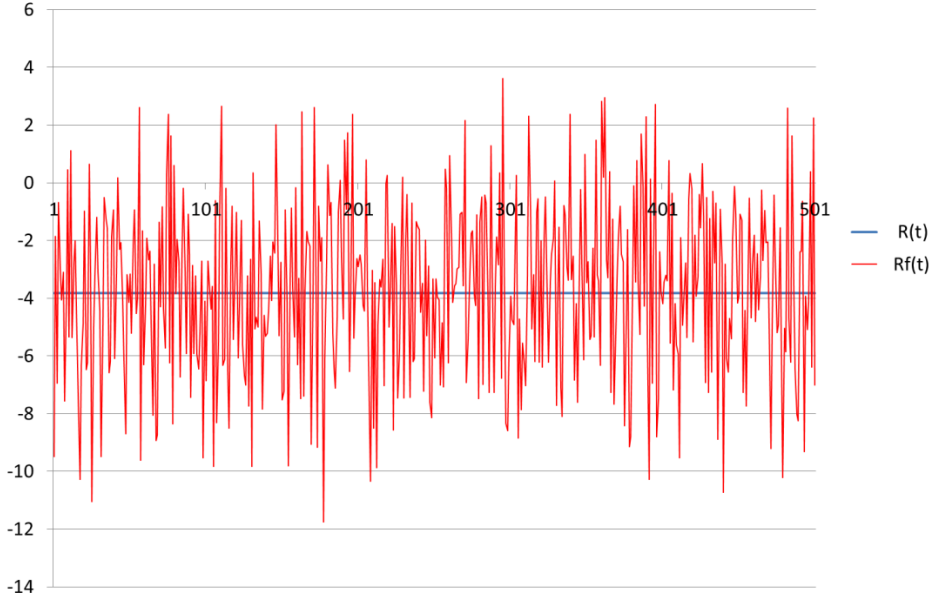

**Figure 7**: Summation of the three components $R_{123}$ (= $\sum_{i=1}^{3} R_i$) with and without noise in the fuzzy interval of (0, 3π/2r); the coupling coefficient $g = 0.2$; initial values as in Figure 3.

Both the fuzzy and the non-fuzzy solutions add up to the same constants, which are shown in Figure 7: $R_{123} = -3.82$ (on average) in this case. Decreasing the length of the fuzzy interval leads to less stochasity in the redundancy and more stability in the reduction of the uncertainty that prevails at the systems level.

## 4. Spectral analysis of empirical data

We have argued above that the mutual redundancy in three or more dimensions can be considered the numerical value of a structural difference between two rotating vectors (*P* and *Q*) in the vector space of university-industry-government relations. By squaring the vectors, this redundancy was expressed in Eq. 2 as the square of the communication field among (three) bilateral communication channels. In the previous section (Eq. 15), these functions were also formulated as composites of sine and cosine functions. Because the difference of the two vectors is constant (Figure 3), the same equation is also valid for the partial redundancies (see Appendix B for the derivation), and therefore one can formulate as follows:

$$R_i = A'_i + B'_i \cos(rt) + D'_i \sin(rt) \tag{18}$$

where $i = 1, 2,$ or $3$ (that is, $U$, $I$, or $G$); $A'_i$, $B'_i$, and $D'_i$ are corresponding constants. Since redundancy is an additive function, each partial redundancy can also be presented in the form similar to Eq. 17 (above). Substitution of Eq. 17 into Eq. 18 leads to:

$$R_i(t) = A'_i + \sum_{k=1}^{N}\left(B'_{ik}\cos(r_k t) + D'_{ik}\sin(r_k t)\right) = S_i(t) \tag{19}$$

We have added "$= S_i(t)$" on the right side to Eq. 19 because this expression is precisely equivalent to the Fourier series if one equates $r_k$ to $kr$. Using Fourier analysis, one is able to decompose a function $R_i(t)$ into a set of $N$ oscillating functions, as follows: $R_i(t) = \sum_{k=1}^{N} R_{ik}$. (In Eq. 19, $r$ is a constant and $k$ denotes the consecutive term in the Fourier series.)

In this context, the coefficients $A'_i$, $B'_{ik}$, $D'_{ik}$ can be considered the frequency spectrum of the TH. One is able to calculate these coefficients and therefore the spectrum when one has time-series data for an empirical TH configuration. In summary, decomposition of the mutual redundancy using Eq. (19) provides us with a spectral analysis of the frequencies in an empirically measured TH system.

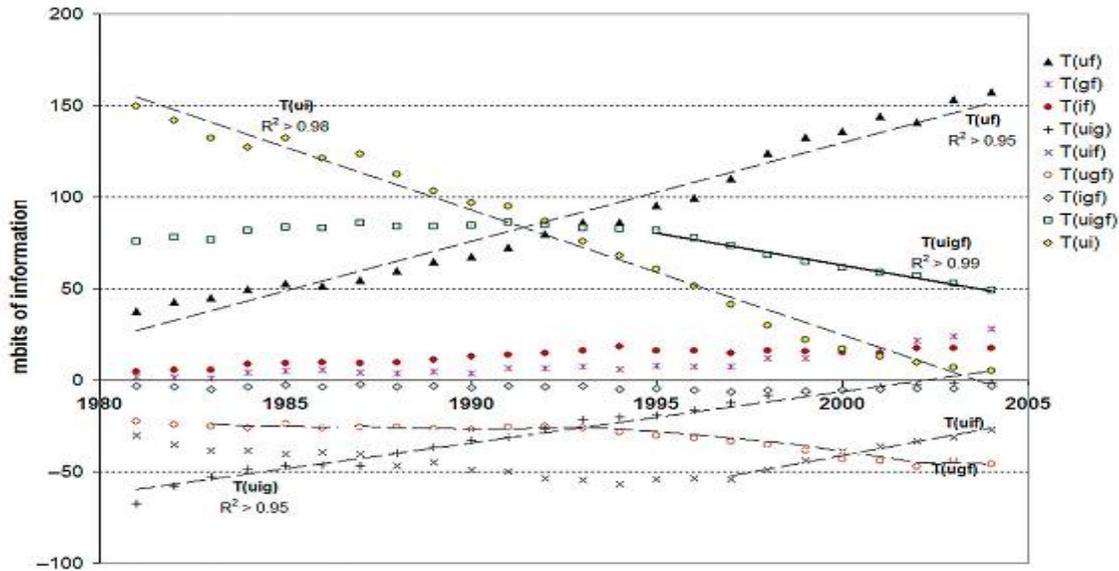

**Figure 8:** The mutual information in two, three and four dimensions among Japanese articles with a university, industrial and governmental address, and international co-authorships. Source: Leydesdorff & Sun (2009, at p. 783).

Let us as an example analyze the time-series in the data provided by Leydesdorff & Sun (2009, at p. 783, Figure 5) using Fourier analysis. Figure 8 first provides the source figure with the time-series of $T_{uig}$, $T_{uif}$, $T_{ugf}$, and $T_{igf}$, that is, the mutual redundancies in three dimensions for different combinations among university, industrial, governmental, or international addresses in Japanese articles during the period 1980-2004. (Leydesdorff & Sun [2009] used the "T" of transmission as a symbol for mutual information, but in this text we have used hitherto the "R" of mutual redundancy. In the case of three dimensions, however, $T_{123} = R_{123}$ [Leydesdorff & Ivanova, in press].) In addition to the three dimensions ($u$, $i$, and $g$), the fourth dimension "$f$" stands for "foreign" addresses (that is, "foreign" from a Japanese perspective).

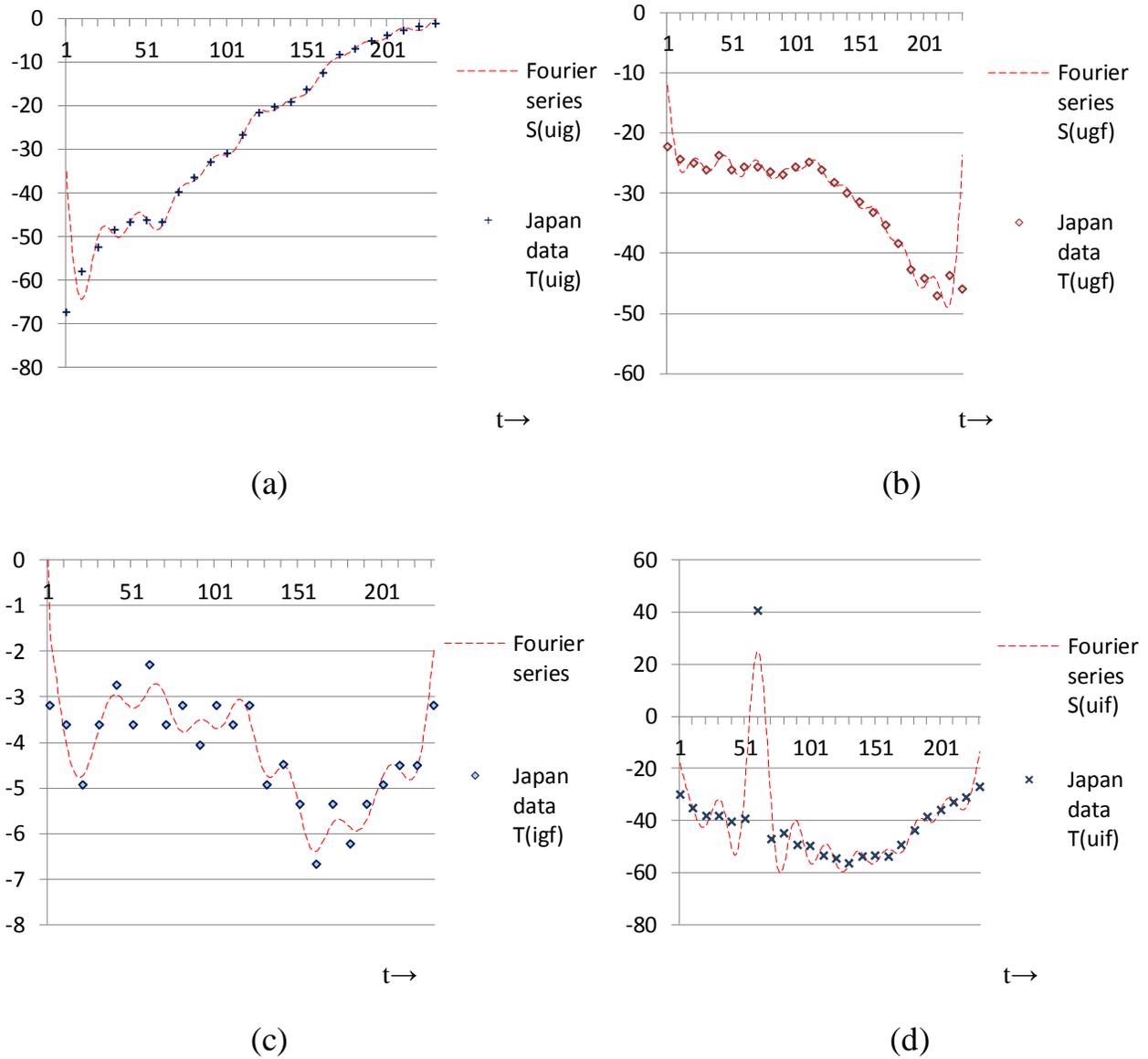

**Figure 9:** Japanese data for mutual redundancy $T_{uig}$ and its approximation by a Fourier series $S_{uig}$ (in panel a); (b) $T_{ugf}$; (c) $T_{igf}$; and (d) $T_{uif}$. The data for the time-series correspond to those in Figure 7.

Figure 9 shows the fit of the four ternary redundancy terms in Figure 7 [$T_{uig}(t)$, $T_{ugf}(t)$, $T_{igf}(t)$, $T_{uif}(t)$] and their approximation by the Fourier series $S(t)$ using $N= 15$ terms for the decomposition.

Different terms in a Fourier series refer to different frequencies in the TH cycles. Each cycle adds a partial redundancy. The relative inputs of the cycles can be measured by considering the vector $B'_{ik}\cos(r_k t)$ and $D'_{ik}\sin(r_k t)$ in Eq.19 as composed by different movements. These vector components can be calculated as the squares of the respective lengths of each vector, as follows:

$$V_{ik} = B'^2_{ik} + D'^2_{ik} \tag{20}$$

The relative inputs of consecutive terms with different frequencies provide us with estimates of the relative values of $V_{ik}$. Figure 10 shows the 15 relative frequencies in the Japanese data using a logarithmic scale.

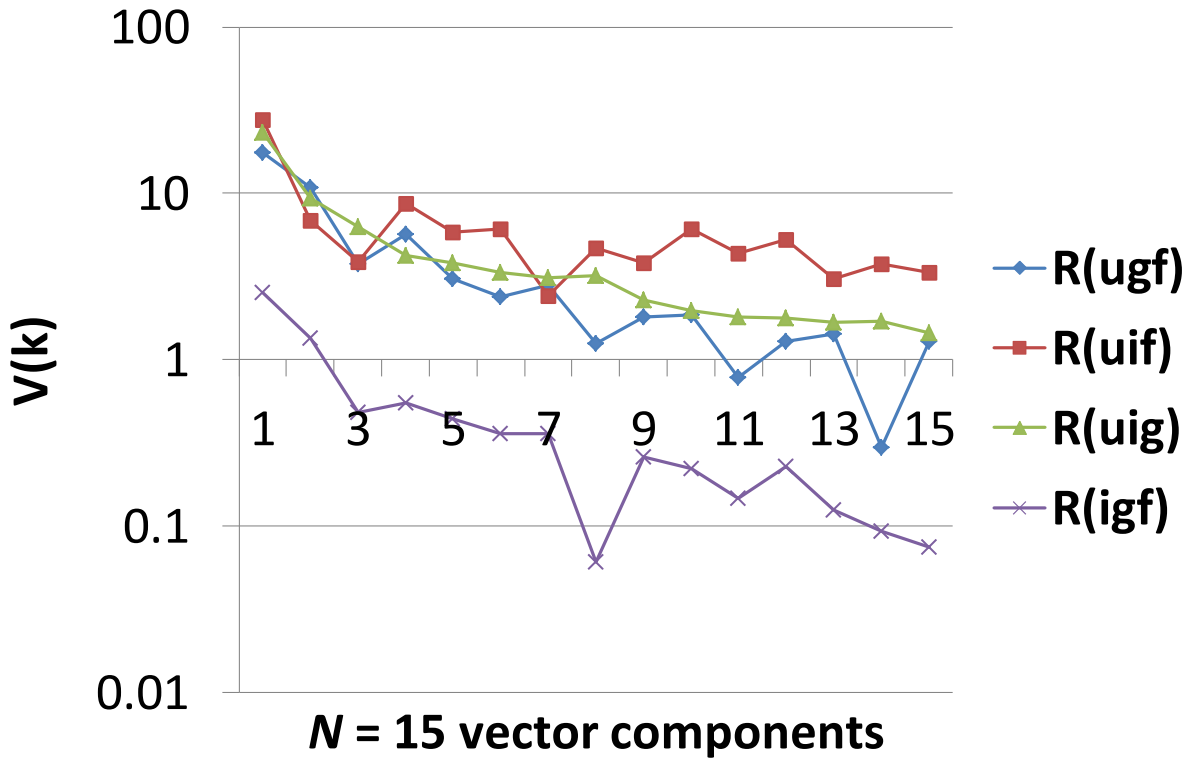

**Figure 10:** Relative frequencies after Fourier analysis of the Japanese data. The *x*-axis denotes consecutive summands in the Fourier series; the *y*-axis shows the relative values of the summands (*k*) using Eq. 20.

Note that the values in Figure 10 are not redundancies, but frequencies of the communications (over time) that generate mutual redundancies such as $R_{uig}$, $R_{uif}$, etc., at each moment of time. The three-dimensional communication fields operate as cycles among the two-dimensional cycles of communication as shown above in Figure 2. The cycles exhibit similar frequencies for the three redundancies in which academia is involved ($R_{uig}$, $R_{uif}$, $R_{ugf}$), but much lower frequencies in the case of $R_{igf}$.

The interpretation is, in this case, trivial: the set is based on institutional co-authorship data in scholarly publications contained in the Web of Science: therefore the non-academic cycle of $R_{igf}$ can be expected to exhibit lower frequencies. However, Figure 9 also shows unequal declines for the various curves: university-industry-foreign relations, for example, initially decline faster to less-frequent cycles, but among the lower frequencies ($k > 8$), these cycles (*uif*) prevail. A further interpretation could elaborate an analogy to theorizing about long (Kondratiev) and short business cycles, but such a theoretical interpretation would lead us beyond the scope of the present study.

In summary, the various redundancies are caused by oscillating terms having different periods. The process is structured, and this structure may also be interpretable in terms of hierarchies among the communications involved, since cycles can be embedded in other cycles. Ivanova & Leydesdorff (in press) showed that the three dimensions can then be expected to shape a fractal manifold.

**Conclusions and summary**

Three or more systems can be differently related as well as differently positioned. The positions are the result of the correlations between the distributions of relations. The relations, however, provide the variation or (Shannon-type) uncertainty that is operationally generated with the arrow of time. Thereafter, the "differences can make a difference" (Bateson, 1973, p. 315) for the systems of reference that are positioned. The positions do not relate, but feedback on the possible relations by structuring a vector space.

The reformulation of mutual information in three (or more) dimensions as mutual redundancy ("overlap") in the positions of the systems as fields by Leydesdorff & Ivanova (in press) enables us to take the next step of considering this redundancy as the result of the operation of a communication field ("overlay" or hyper-cycle) that can emerge auto-catalytically on top of the three (bilateral) communication channels. The operation of this field can be measured (as a scalar) in terms of mutual redundancy ($R_{123}$), but the interpretation of this measure in terms of relative positions enables us to simulate the system in the vector space.

The three dimensions (university, industry, government) provide selection environments for each other, but in a "double contingency" (Parsons, 1951): (1) variation is generated in the exchanges and can be measured as Shannon entropy, and (2) the selections operate in terms of these relations, but are structure-determined, and can be modeled using Lotka-Volterra equations. Because this structure is dually layered—in terms of Shannon-type communication relations and meaning provided to these exchanges—the equations for three selections operating upon one another as components of two vectors (*P* and *Q*) could be solved analytically (Appendix A).

This result enables us to simulate the various components of $P_i$ and $Q_i$, and then also $R_i$ for the three dimensions. Because *R* is a signed information measure (Yeung, 2009: 59f.) and an additive scalar, it follows that $R = \sum_{i=1}^{3} R_i \sim \sum_{i=1}^{3}(P_i^2 - Q_i^2)$. Thus, one can bring the vector-perspective in accordance with the discrete information-theoretical one of the measurement.

Not incidentally (Appendix B), it could thereafter be shown that these vectors can be formulated as Fourier series in terms of sine and cosine functions. These smooth functions are not directly affected by the noise that operates in terms of phase differences between the functions in fuzzy intervals. The spline-functions, however, can be decomposed using Fourier analysis. As an example, we used data previously published for the Japanese publication system. Whereas the various frequencies operating in the different loops in this data could be shown, an interpretation of this specific set would go beyond the framework of this study and is perhaps meaningful only on the basis of a more globally selected sample.

Our argument here is that a relation can be constructed between the mathematical simulation and the empirical measurement because of a number of conditions in the TH model that make specific systems of equations solvable:

1. A TH system can self-generate an overlay structure that can be expected to feedback on the underlying (i.e., bilateral) channels and thus be expected to shape a fractal structure (Ivanova & Leydesdorff, in press). Each TH thus contains a set of other TH-structures, such as national innovation systems containing regional innovation systems, but also sectorial ones that extend across regions. The system is fractal and nested in different directions (Leydesdorff, 2006).

2. The double-layering of relations and positions makes the Lotka-Volterra equations modeling three selection environments that operate upon one another solvable because the variation terms are analytically distinct from the selection mechanisms. (The parameters for the variation can be set equal to zero when modeling the interactions among selecting structures because the variation can be considered as stochastic in a first approximation, and therefore expected to average out.)

3. The systems of equations that are thus generated can be solved in terms of sine and cosine functions (Kamke, 1971). This solution could first be used for the simulations, and secondly for the Fourier decomposition of frequencies in the empirical time-series data.

The confluence of these steps into a single and coherent model is convenient for bringing the measurement and simulation of TH configurations into the longer-term perspective of specifying expectations of how the different TH-components in a configuration can be expected to influence uncertainties in the complex system(s) under study. For example, in Figure 9 we saw that internationally co-authored publications exhibit other communication frequencies than university-industry publications which are co-authored with national government agencies (in the case of Japan). The contribution to the redundancy of each TH partner ($R_i$) could also be specified.

The model we have developed above is in principle appliable to any "triple contingency" (Strydom, 1999; cf. Burt, 2000) and thus may help to solve the long-standing problem of how to model and measure the sharing of meaning among reflexive agencies in interhuman communications. In this study, we have used the TH metaphor as a ladder, but in a next study we should throw this ladder away and take up the problem that Shannon's (1948) mathematical theory of communication cannot address: the communication of meaning or meaningful information—as Shannon himself emphasized (cf. Leydesdorff & Franse, 2009).[7]

---
[7] "Frequently the messages have *meaning*; that is they refer to or are correlated according to some system with certain physical or conceptual entities. These semantic aspects of communication are irrelevant to the engineering problem." (Shannon, 1948, at p. 379).

**Appendix A**. *Communication of information and meaning in a TH dynamic symmetry model.*

In a TH dynamic symmetry model, communication field $W$ is described by the following equation:

$$\partial^j W_{ij} = -gW^j \times W_{ij} \tag{A.1}$$

Here $i,j = 0,1$; $\partial_0 = \partial_t$; $\partial_1 = -\partial_x$; $\partial^0 = \partial_t$; $\partial^1 = \partial_x$; summing on repeating indexes $j$ is implied. Eq. (A.1) can be rewritten in a more explicit form:

$$\begin{aligned}\partial^1 W_{01} &= -gW^1 \times W_{01} \\ \partial^0 W_{10} &= -gW^0 \times W_{10}\end{aligned} \tag{A.2}$$

$W_{ij}$ is defined by the formula:

$$W_{ij} = \partial_i W_j - \partial_j W_i + gW_i \times W_j \tag{A.3}$$

$W_i$, $W_j$ are three-component vectors in TH internal symmetry space, and two component vectors in $(x, t)$ space. We can define these as: $W_i = (P,Q)$; $W^i = (-P,Q)$; $P, Q$ are three-component vectors in the case of university-industry-government (TH) relations. Eq. (A.3) can be expressed in components:

$$\begin{cases} W_{01} = Q_t + P_x + g(P \times Q) \\ W_{10} = -Q_t - P_x - g(P \times Q) \end{cases} \tag{A.4}$$

and substituting it into (A.2) produces:

$$\begin{cases} P_{xx} + Q_{tx} = -g(P_x \times Q) - g(P \times Q_x) - gQ \times \left[P_x + Q_t + g(P \times Q)\right] \\ -P_{xt} - Q_{tt} = g(P_t \times Q) + g(P \times Q_t) - gP \times \left[P_x + Q_t + g(P \times Q)\right] \end{cases} \tag{A.5}$$

Setting: $P = P(\xi)$; $Q = Q(\xi)$; $\xi = x - t$, so that $P_x = -P_t = P_\xi$; $Q_x = -Q_t = Q_\xi$, summing first and second equations in (A.5), and assuming that: $Q \times Q_t = gQ \times (P \times Q)$; $P \times P_x = gP \times (P \times Q)$, we can rewrite Eq. (A.5) in the form:

$$2P_{\xi\xi} - 2Q_{\xi\xi} = 2(P \times Q) \times (g^2 Q + g^2 P) - g(P \times Q_\xi) - g(Q \times P_\xi) \tag{A.6}$$

This can be interpreted as describing two fields: *P*, *Q* propagating in a positive direction of the *x* axes.

We can show that Eq.(A.6) can be coupled to the equation describing these two fields *P*, *Q* propagating in a negative direction of the *x* axes (or propagating in a positive direction but reversed in time). Setting:

$$\begin{cases} P_t = -2\beta(P \times Q) \\ Q_t = +2\delta(Q \times P) \end{cases} \quad (A.7)$$

where *P*, *Q* are vectors: $P = \begin{pmatrix} P_1 \\ P_2 \\ P_3 \end{pmatrix}$; $Q = \begin{pmatrix} Q_1 \\ Q_2 \\ Q_3 \end{pmatrix}$, and the right sides of (A.7) are cross-products.

Alongside the "time" version of equation (A.7) can be introduced the "space" version of this equation:

$$\begin{cases} 2P_x = -\beta(P \times Q) \\ 2Q_x = +\delta(Q \times P) \end{cases} \quad (A.8)$$

Each pair of equations (A.7) and (A.8) describes the evolution of functions *P* and *Q* in time and space. We can obtain a combined space-time equation. Differentiating the first and second equations of the system (A.8) by *t* and taking into consideration (A.7) produces:

$$\begin{cases} 2P_{xt} = +2\beta^2(P \times Q) \times Q - \beta(P \times Q_t) \\ 2Q_{xt} = -2\delta^2(P \times Q) \times P + \delta(Q \times P_t) \end{cases} \quad (A.9)$$

Setting further: $P = P(\xi)$; $Q = Q(\xi)$; $\xi = x + t$, so that $P_x = P_t = P_\xi$; $Q_x = Q_t = Q_\xi$, and subtracting in (A.9) the second equation from the first, we obtain:

$$2P_{\xi\xi} - 2Q_{\xi\xi} = 2(P \times Q) \times (\beta^2 Q + \delta^2 P) - \beta(P \times Q_\xi) - \delta(Q \times P_\xi) \quad (A.10)$$

Eq.(A.10) coincides in full with Eq. (A.6) if we set $\beta = \delta = g$, but refers to solutions which evolve reversed in time.

**Appendix B**. *Analytical solution for a TH communication field.*

Defining the functions

$$f_1 = 2g(P_2 Q_3 - P_3 Q_2)$$
$$f_2 = 2g(P_3 Q_1 - P_1 Q_3) \quad \text{(B.1)}$$
$$f_3 = 2g(P_1 Q_2 - P_2 Q_1)$$

We can write a system of equations for a TH communication field (coupling constant $2g$ in the right side of systems (2), (3) not mentioned)

$$P_{1t} = -f_1 \qquad Q_{1t} = f_1$$
$$P_{2t} = -f_2 \qquad Q_{2t} = f_2 \quad \text{(B.2)}$$
$$P_{3t} = -f_3 \qquad Q_{3t} = f_3$$

from system (2) we obtain:

$$P_{1t} + Q_{1t} = 0 \qquad\qquad P_1 + Q_1 = \alpha$$
$$P_{2t} + Q_{2t} = 0 \quad \text{or otherwise:} \quad P_2 + Q_2 = \beta \quad \text{(B.3)}$$
$$P_{3t} + Q_{3t} = 0 \qquad\qquad P_3 + Q_3 = \gamma$$

Expressing Q in terms of P from system (B.3) and substituting the result into system (B.2), we get:

$$P_{1t} = cP_2 - bP_3$$
$$P_{2t} = aP_3 - cP_1 \quad \text{(B.4)}$$
$$P_{3t} = bP_1 - aP_2$$

Here: $a = -2g\alpha$; $b = -2g\beta$; $c = -2g\gamma$. System (B.4) can be solved analytically and has a well-known solution (Kamke, 1971)

$$P_1 = aC_0 + rC_1 \cos(rt) + (cC_2 - bC_3)\sin(rt)$$
$$P_2 = bC_0 + rC_2 \cos(rt) + (aC_3 - cC_1)\sin(rt) \quad \text{(B.5)}$$
$$P_3 = cC_0 + rC_3 \cos(rt) + (bC_1 - aC_2)\sin(rt)$$

Here: $r^2 = a^2 + b^2 + c^2$; $\quad aC_1 + bC_2 + cC_3 = 0$ .

From the equation:

$$aP_{1t} + bP_{2t} + cP_{3t} = 0 \qquad (B.6)$$

we get after integration:

$$aP_1 + bP_2 + cP_3 = C_4 \qquad (B.7)$$

and from the equation:

$$P_1 P_{1t} + P_2 P_{2t} + P_3 P_{3t} = 0 \qquad (B.8)$$

we also get:

$$P_1^2 + P_1^2 + P_1^2 = C_5 \qquad (B.9)$$

This means that integral curves are situated simultaneously on the sphere surface. The same equation holds for $Q$:

$$Q_1^2 + Q_1^2 + Q_1^2 = C_6 \qquad (B.10)$$

From (5) we get:

$$a^2 C_0 + b^2 C_0 + c^2 C_0 = aP_1(0) + bP_2(0) + cP_3(0) + r(aC_1 + bC_2 + cC_3)$$

So that:

$$C_0 = \left(aP_1(0) + bP_2(0) + cP_3(0)\right)/r^2 \qquad (B.11)$$

Constants $C_1$, $C_2$, $C_3$ can be calculated from system (5):

$$\begin{aligned} C_1 &= \left(P_1(0) - aC_0\right)/r \\ C_2 &= \left(P_2(0) - bC_0\right)/r \\ C_3 &= \left(P_3(0) - cC_0\right)/r \end{aligned} \qquad (B.12)$$